\documentclass[aps,prb,twocolumn,superscriptaddress,showpacs]{revtex4-1}

\usepackage{amsmath}
\usepackage{graphicx}

\usepackage{color}

\begin{document}

\title{Magnetic field control of intersubband polaritons in narrow-gap semiconductors}

\author{Giovanni Pizzi}
\affiliation{Scuola Normale Superiore, Piazza dei Cavalieri 7, I-56126
  Pisa, Italy}
\affiliation{NEST, Istituto Nanoscienze-CNR, Piazza San Silvestro 12, I-56127 Pisa, Italy}
\affiliation{Laboratoire Pierre Aigrain, Ecole Normale Sup\'erieure, CNRS UMR 8551, Universit\'e P. et M. Curie, Universit\'e Paris Diderot, 24 rue Lhomond, F-75005 Paris, France}
\author{Francesca Carosella}
\author{G\'erald Bastard}
\author{Robson Ferreira}
\email{robson.ferreira@lpa.ens.fr}
\affiliation{Laboratoire Pierre Aigrain, Ecole Normale Sup\'erieure, CNRS UMR 8551, Universit\'e P. et M. Curie, Universit\'e Paris Diderot, 24 rue Lhomond, F-75005 Paris, France}

\date{\today}

\begin{abstract}
We investigate theoretically the polariton coupling between the light confined in a
planar cavity and the intersubband transitions of a two-dimensional
electron gas confined in semiconductor quantum wells in the presence
of a vertical magnetic field.
We show that in heterostructures made of non-parabolic
semiconductors, the polaritons do not fit a two-level
problem, since the cavity photons couple to a non-degenerate
ensemble of intersubband transitions. As a consequence, the stationary polariton eigenstates
become very sensitive to the vertical magnetic field, which thus plays the
role of an external parameter that controls the
regime of light--matter interactions. At intermediate field strength
we predict that the magneto-polaritons have energy dispersions ideally
suited to parametric amplification. 
\end{abstract}


\maketitle

\section{Introduction}
Intersubband (ISB) polaritons are mixed states formed by the strong
coupling of the light within a microcavity and the intersubband
transitions of electrons confined in a semiconductor quantum well (QW)
embedded in the cavity.  
Since the first experimental demonstration in 2003\cite{Dini:2003}
with a GaAs/AlGaAs multiple quantum well (MQW) structure,
intense research efforts have been devoted to the study of ISB
polaritons. With this kind of polaritons the light--matter coupling
can reach very large values\cite{Dupont:2007,Anappara:2007} becoming comparable to (or even larger than)
the bare frequency of the cavity and of the ISB excitations.
In this
ultrastrong coupling regime, interesting 
quantum effects appear.\cite{Ciuti:2005,DeLiberato:2007,DeLiberato:2009a,Nataf:2010}
Moreover, since the coupling strength is proportional to the square root of the number of
electrons, it can be controlled by electrical
gating.\cite{Anappara:2005,Guenter:2009}
Beside the observation of the strong coupling regime by means
  of reflectance spectroscopy as in the first experiments, and of photovoltaic
  measurements,\cite{Sapienza:2007} also the electrical injection of
  cavity polaritons and their electroluminescence is being studied with considerable effort.\cite{Colombelli:2005,Sapienza:2008,Todorov:2008,Jouy:2010}
 Moreover, the coupling of the ISB transition with a surface plasmon
 supported by a metal grating has been
 demonstrated.\cite{Zanotto:2010}
In the effort of reaching larger light--matter couplings
toward the ultrastrong coupling regime, other 
materials beside GaAs/AlGaAs have been considered,
like for instance InAs/AlSb
MQWs. Also the smaller effective mass of
InAs with respect to GaAs
($m^*_{\text{InAs}} /m^*_{\text{GaAs}}=0.39$) implies a stronger coupling.\cite{Anappara:2007}
At zero magnetic field, the polaritons can be simply and effectively
described by a two-level problem,\cite{Ciuti:2005} where the first
level is the cavity mode with energy $E_\text{cav}$, and the
second level is the ISB transition with energy $E_{21}$ between the first (ground) and the
second (excited) subband; the coupling is quantified by the Rabi
frequency $\Omega_R$, where $2\hbar\Omega_R$ gives the splitting of the
upper and lower polariton branches at the resonance $E_\text{cav}=E_{21}$.

When a magnetic field $B$ is applied along the QW growth axis $\hat
z$, neither the energies nor the strength of the ISB--cavity coupling
are altered; thus, \emph{a fortiori}, the two-level description
of the polariton levels remains valid, if we still
  focus on transitions between the Landau levels belonging to
  different subbands. A theoretical study on the possibility
  of obtaining ultrastrong magneto-polaritons couplings
  exploiting transitions between Landau levels in the same subband is
  reported in Ref.~\onlinecite{Hagenmuller:2010}.
Actually, the aforementioned insensitivity to a vertical magnetic field is exact only for
parabolic-band materials. It remains a
very good approximation 
for GaAs-based heterostructures, since GaAs shows very little
non-parabolicity.  
On the contrary, as we show below, in narrow-gap semiconductors like InAs or InSb, the
band non-parabolicity effects cannot be
disregarded in the calculation of the polaritonic states.

In this work, we demonstrate that in the non-parabolic case the ISB
polaritons cannot be simply described in terms of two levels.
Instead,
the cavity photons couple to a non-degenerate ensemble of ISB
transitions, giving rise to a complex evolution of the polariton
dispersion for increasing $B$. We shall show that three different
coupling regimes exist as a function of the intensity of the magnetic field.
To this end, we consider a InAs/AlSb
MQW heterostructure grown along the $\hat z$ axis. This choice is
motivated by the experimental observation of ISB polaritons in this
system,\cite{Anappara:2007}  as well by the
significant band non-parabolicity of InAs.
Band parameters and band
offsets are taken from Ref.~\onlinecite{Vurgaftman:2001}. We consider a 
cavity with
effective thickness $L_z$ so that for the lowest mode $k_z=\pi/L_z$ holds.
The photon energy is given by
$E_\text{cav}=(\hbar c/\sqrt{\varepsilon_\infty})
\sqrt{k_{\parallel}^2+k_z^2}$, where $k_{\parallel}$ is the in-plane $\mathbf{k}$
vector and $\varepsilon_\infty$ is the dielectric constant of the
material embedded in the cavity. Due to the usual ISB
selection rules, we consider only light which is TM polarized
(i.e.\@ with a component of the electric field along the growth axis
$\hat z$).

\section{Results and discussion}
\begin{figure}[t]
\includegraphics{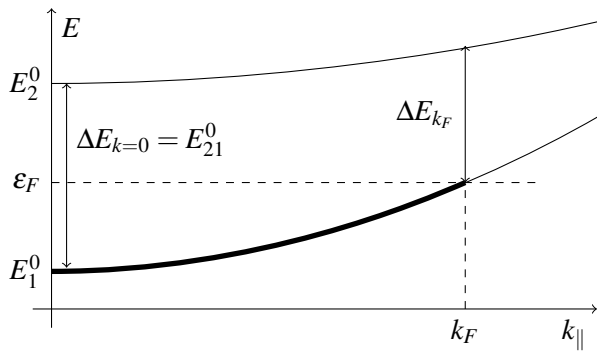}%
\caption{Subbands of a non-parabolic material at $B=0$
 (not to scale). Since the two subbands have different masses
  and thus different curvatures,
  the transition energy depends on the in-plane wavevector. We calculate: 
  $\varepsilon_F-E_1^0\approx 27$~meV and $\Delta E_c=\Delta E_{k=0}-\Delta
  E_{k_F}\approx 11$~meV (while $E^0_{21}=310$~meV).\label{fig:subbands-zero-B}}
\end{figure}

The non-parabolicity effect can be 
described in a QW\cite{Bastard:1991} by an
effective mass for the in-plane motion $m^*_n$, depending on the
subband index $n$. 
We set the QW width to 6.6~nm and calculate that the ISB transition
energy at zero magnetic field is $E^0_{21}=310$~meV, and the effective
masses for the first and second 
subbands are  $m^*_1=1.68\,m_{\Gamma_6}$ and 
$m^*_2=2.85\,m_{\Gamma_6}$, where $m_{\Gamma_6}=0.026\,m_0$ is the InAs bulk
effective mass. For the simulation we choose the cavity
mode coupled to $n_{QW}=5$~QWs, each with an electron density
$n_{2D}=5\cdot10^{11}$~cm$^{-2}$ in its first subband ($n=1$; all
calculations are performed at zero temperature). At $B=0$, the
ensemble of ISB transitions forms a
finite-width continuum. As depicted in
Fig.~\ref{fig:subbands-zero-B}, this is due to the fact that 
the two subbands have different curvatures (since
$m^*_1 \neq m^*_2$). Therefore, the ISB
transition energy depends on the in-plane wavevector $k_\parallel$,
reaching its minimum value at the Fermi wavevector $k_\parallel =
k_F$. The width of the continuum is given by $\Delta E_c = \pi\hbar^2 n_{2D}(1/m_1^*-1/m_2^*)$. 

With the application of a magnetic field $B$ along the
growth axis, each subband splits into a set of discrete Landau
levels (LLs) with approximate energies 
\begin{equation}
\label{eq:Landau-Level-Non-Parabolic}
E_{n,j}(B)=E_{n}^0 + \left(j+\frac 1 2\right)\frac{\hbar e B}{m^*_{n}},
\end{equation}
where $n$ is the subband index, $j=0,1,\ldots$ the LL index,
and $E_n^0$ is the subband edge energy at $B=0$. 
In Eq.~\eqref{eq:Landau-Level-Non-Parabolic}
we have assumed that the effective mass depends mainly on the subband
index $n$ and not on the LL index $j$, which is a reasonable assumption
as far as the LL separation is much smaller
than the intersubband transition energy (this was checked for all
relevant values of the $B$ field).
Note that, here and
in the following, we do not consider explicitly the Zeeman spin splitting of
the LLs, since the ISB transitions are spin-conserving.

Within the electric dipole approximation, the ISB transitions verify
$\Delta j=0$.
The transition energy $\Delta E_j(B)=E_{2,j}-E_{1,j}$ for electrons in
the $j$-th LL is then given by 
\begin{equation}
\label{eq:Transition-Energy}
\Delta E_j(B)=E_{21}^0 + \left(j+\frac 1 2\right)\hbar e
B\left(\frac{1}{m^*_2}-\frac{1}{m^*_1}\right).
\end{equation}

We note that in the parabolic case, since $m^*_1=m^*_2=m^*$, we obtain
as expected that $\Delta E_j(B)=E_{21}^0$ does not depend on the
magnetic field: therefore all transitions for the different LLs are
degenerate at the same energy $E_{21}^0$, and we can safely apply the
same two-level formalism as used at $B=0$.

\begin{figure}[t]
\includegraphics{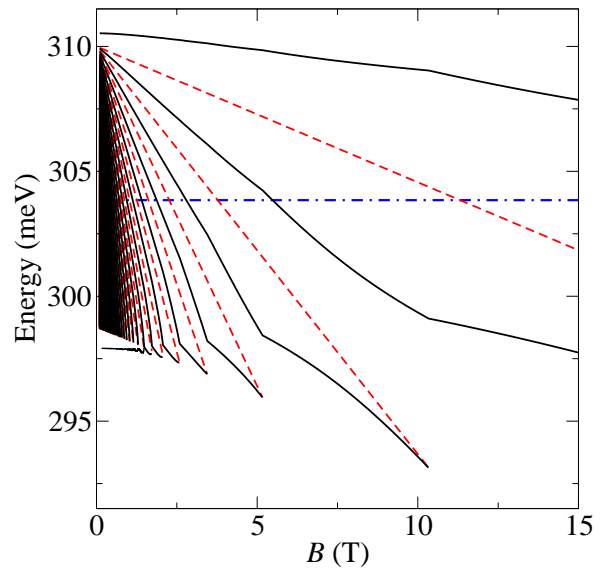}%
\caption{(Color online) Polariton branches at fixed in-plane vector
  $k_{\parallel}$ versus the
  magnetic field $B$ for a InAs/AlSb
  MQW structure embedded in a microcavity (black solid lines). 
  Blue dotted--dashed line: bare cavity mode energy
  $E_{cav}$.  Red dashed lines:
 ISB transition energies $\Delta E_j(B)$, plotted
  only in the $B$ range in which the corresponding LLs in the
  ground subband are not empty. For the parameters, 
  see text.\label{fig:transitions-B}} 
\end{figure}

In the non-parabolic case we have instead
an ensemble of transitions at different $B$-dependent energies.
In particular, since  
$m_2^*>m_1^*$, $\Delta E_j(B)$ decreases with $B$ for all $j$ values, as shown by
red dashed lines in Fig.~\ref{fig:transitions-B}. The number of
active ISB transitions is given by the number of filled LLs in the
ground subband $n=1$, and thus also depends on $B$. In
Fig.~\ref{fig:transitions-B}, each $\Delta E_j$
transition energy is in fact plotted versus $B$ only in the $B$ range for which
the $E_{n=1,j}$ level is not empty, i.e.\@
for $B<B_j=\frac {\pi \hbar n_{2D}}{e}\cdot \frac {1}{j}$ for $j\geq 1$ (while the
$j=0$ LL is always filled).

In order to have a significant coupling with more than one ISB level,
we choose a cavity geometry with a cavity mode energy
$E_{\text{cav}}<E_{21}^0$ (blue dotted--dashed line of
Fig.~\ref{fig:transitions-B}), so that in absence of coupling the
photon energy
crosses the bare ISB transition energies.
In particular, in Fig.~\ref{fig:transitions-B} we choose an
effective cavity thickness $L_z=1.5$~$\mu$m and we set $\varepsilon_\infty=12.32$
for the dielectric constant of the InAs cavity. The in-plane
wavevector is fixed to $k_\parallel = 4.99$~$\mu$m$^{-1}$, corresponding to an
angle of propagation inside the cavity of $\theta=67.2^\circ$ with
respect to the $\hat z$ axis.

For the calculation of the polaritons, we note that each allowed transition
channel $j$ is independent of the others and occurs at a different
energy $\Delta E_j$ (for $B\neq 0$). We thus describe the
polariton eigenstates (for a given in-plane $k_\parallel$ vector) as a
linear combination of the state $|a\rangle$ with one photon in the cavity mode and no
ISB excitations, and the set of states $|b_j\rangle$ with one ISB
excitation associated to a given LL $j$ and no
photons in the cavity. The
coupling between the $|a\rangle$ and $|b_j\rangle$ states is then given by 
\begin{equation}
\label{eq:Coupling-Frequencies}
\Omega_j = \tilde\Omega \sqrt{\frac{n_j}{n_{2D}}},
\end{equation}
where $n_j$ is the population of the $j-$th LL in the ground
subband (so that $\sum_j n_j=n_{2D}$). The frequency $\tilde\Omega$
is calculated in a way similar to the $B=0$ case:\cite{Ciuti:2005}
$$
\hbar \tilde\Omega =
\sqrt{\frac{e^{2}\hbar^{2}n_{2D}n_{QW}E_{21}^0f_{21}\sin^{2}\theta
  }{2m_0\varepsilon_{0}\varepsilon_{\infty}E_{cav}L_{z}}} 
$$
where $f_{21}$ is the oscillator strength, which has been defined and calculated
taking into account the non-parabolicity as described
in Ref.~\onlinecite{Sirtori:1994}: $f_{12} = 0.79\,
m_0/m_{\Gamma_6}$ for our parameters.
The calculated polariton branches are represented by black solid lines in
Fig.~\ref{fig:transitions-B}. Note that, for a given value of $B$,
we have included in the calculation only the $|b_j\rangle$
states originating from non-empty LLs.

The system parameters have been chosen in order to achieve a significant
coupling between the cavity mode $|a\rangle$ and more than one ISB
transition level $|b_j\rangle$. This can be achieved only if the coupling energy
$\hbar \tilde\Omega$  is of the order of the typical deviation of the ISB
transition energies with respect to $E_{21}^0$. In fact, if the
coupling is much larger than the energy separation 
between the different ISB transitions, the latter ones
behave essentially as a single degenerate level for what concerns the
coupling with the cavity photons, and we  then
recover the ordinary two-level regime (not shown).  In InAs/AlSb
heterostructures, the ISB transition energy deviation is typically of the order
of 10--15~meV at $B\approx 10$ T (see Fig.~\ref{fig:transitions-B}).
For the parameters used in Fig.~\ref{fig:transitions-B}, the coupling
$\hbar \tilde\Omega$ is about 5~meV.
We notice moreover that
the effects of the squared vector potential can be safely neglected in
our structure,
since the $A^2$ correction is of the order\cite{Ciuti:2005} of
$\hbar^2\tilde\Omega^2/E^0_{21}$,
which is much smaller than
cavity energy $E_{cav}$ (in our case $\hbar \tilde\Omega \ll
E_{cav}\approx E^0_{21}$).

From Fig.~\ref{fig:transitions-B} it is apparent that the polariton
levels display a complex evolution as a function of
the magnetic field. We distinguish three field regions.
For large $B$ fields, where only the $j=0$ LL is
filled, we recover the two-level correspondence, valid also for parabolic
materials. As $B$ decreases, however, more LLs start to be filled
 and as a consequence, more ISB transitions couple to the light. 
At $B=B_j$ ($j\geq 1,2,\ldots$) a new state appears, which however has a
zero coupling at
this precise value of the magnetic field, since the corresponding LL is empty. Decreasing $B$,
its population increases (while the populations of lower LLs
decrease), so that the coupling is spread between the 
levels.
Finally, in the $B\to 0$ limit, we end up with a bare cavity mode coupled
to (and placed inside) a finite-width continuum of ISB transitions. 
As it is well known,\cite{Cohen:1998} the resulting eigenstates
depend on the ratio between the continuum width and the coupling
strength. 
Since in our case $\Delta E_c \approx 2\hbar \tilde\Omega$, two
polariton states appear near each side of the bare ISB 
continuum limits (see Fig.~\ref{fig:transitions-B} and discussion below).

\begin{figure}[t]
\includegraphics{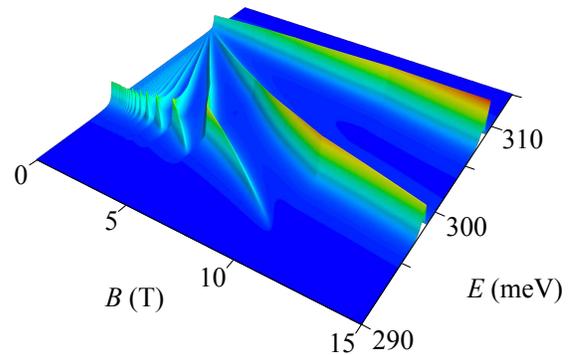}%
\caption{(Color online) Squared modulus of the ``light'' part of the polariton
  eigenvectors. Same parameters as in Fig.~\ref{fig:transitions-B}.
\label{fig:eigenvectors}}
\end{figure}

From the above discussion, we see that the magnetic field assumes the role of
a real external 
control parameter, which can be used to tune the regime of
light--matter interactions. 
To illustrate more clearly this point, we
study also the eigenvector components of the polariton eigenstates.
In Fig.~\ref{fig:eigenvectors} we show the squared modulus of the ``light''
part of the polariton eigenvectors, i.e.\@ the component of the
eigenvectors associated to the state $|a\rangle$. The
magnitude of this component displays the three different
regimes mentioned above. At large $B$ fields, we clearly identify the two
polaritons resulting from the strong coupling between light and the
$j=0$ ISB transitions. In the opposite $B$ regime, i.e.\@ for small
$B$ values, we see that the light component is mainly concentrated on
the two extremal polariton branches. All other
polaritons have a significantly smaller light component, and thus this
regime resembles a two-level regime. Note however that
all states have an influence on the overall coupling
also in the $B\to 0$ limit, and therefore they cannot be disregarded in the
calculation of the polariton coupling. 
Finally, there is a third regime for intermediate values of the
magnetic field: in this case, the light is coupled with a discrete
set of ISB states, and the resulting polariton branches have a similar
magnitude of the light component. 

\begin{figure}[t]
\includegraphics{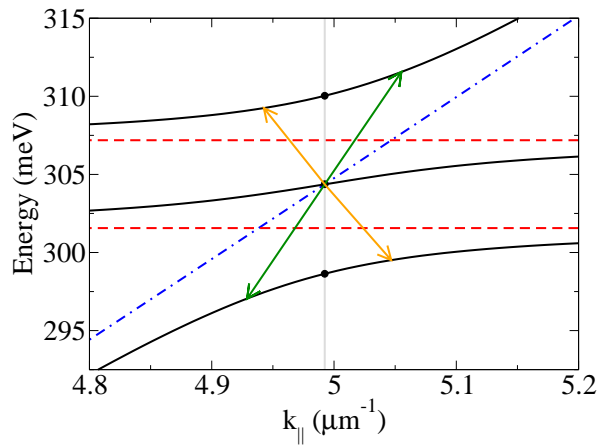}%
\caption{(Color online) Polariton branches as a function of the in-plane
  wavevector (black solid lines) for 
  $B=B_2=5.17$~T.  The 
  blue dotted--dashed line (red dashed lines) represent the cavity
  mode ($j=0$ and $j=1$ ISB transitions) in absence of
  coupling. The vertical line
  indicates the $k_{\parallel}=k_{\parallel R}$ value (see text) at which Fig.~\ref{fig:transitions-B} is
  calculated. Orange and green arrows: sets of entangled states in an
  optical parametric oscillator phenomenon (see text). Same
  parameters of Fig.~\ref{fig:transitions-B}. 
\label{fig:transitions-q}}
\end{figure}

To discuss more in detail the intermediate regime, we focus on a magnetic field
$B=B_2=5.17$~T, which corresponds 
to a complete filling of the $j=0$ and $j=1$ LLs. Since the two
LLs have the same population, 
Eq.~\eqref{eq:Coupling-Frequencies} implies that the respective
states $|b_0\rangle$ and $|b_1\rangle$ couple to the light with the same
strength $\hbar \tilde \Omega/\sqrt{2}$.
As a consequence of this and of the relative energy position of the
uncoupled levels, the three resulting polaritons states have
a similar magnitude of the light component, as it can also be deduced from
Fig.~\ref{fig:eigenvectors}.
The dispersion of these three polaritons (at $B=B_2$) as a function of the in-plane
wavevector $k_\parallel$ is shown in
Fig.~\ref{fig:transitions-q}; the uncoupled cavity mode frequency and
intersubband transitions are also shown with blue dotted--dashed and
red dashed lines, respectively. The vertical line indicates
the value of $k_\parallel$ used in
Fig.~\ref{fig:transitions-B}. 

The magnetic field control of the ISB polaritons might be observable
in an optical experiment.  For very strong $B$ fields (not discussed
here) the cavity mode is energetically isolated, well above all the
ISB levels, so that any spectrum probing the light component of the
system eigenstates should reveal a single intense peak at the bare
cavity energy.
Decreasing to the high fields of Figs.~\ref{fig:transitions-B} and
\ref{fig:eigenvectors} ($B\approx 15$~T), the 
optical spectrum is expected to display instead two peaks,
characteristic of the strong ISB ($j=0$)--cavity coupling.  The spectrum
evolves then into three peaks of comparable intensities when
decreasing the field to $B\approx B_2$. More peaks are expected
to emerge when we further decrease $B$, if the broadening is
small enough to allow to resolve them; the central
peaks should decrease in amplitude as $B$ is further decreased, to the
benefit of the two main lines at  $B=0$. 

Let us finally discuss an interesting aspect of the intermediate field
region, resulting from the existence of multiple polariton lines.  In
Fig.~\ref{fig:transitions-q},  for
$k_\parallel=k_{\parallel R}$ (vertical gray line) the cavity mode lays exactly
at mid-distance from the two $j=0$ and $j=1$ ISB transitions.  Moreover,
since the bare cavity dispersion is to a good approximation a linear
function of $k_\parallel$ around $k_{\parallel R}$, it can be easily shown that the resulting
polariton dispersions $E_m(k_\parallel)$ (with $m=1,2,3$ for the three branches)
have the following interesting ``mirror'' property: they fulfill
$E_3(k_{\parallel R}+k)+E_1(k_{\parallel R}-k) = 2 E_2(k_{\parallel
  R})$, with $k$ a small deviation from the
resonance wavevector.  Two of such sets of three-polariton states are
pictured by green and orange arrows in Fig.~\ref{fig:transitions-q}.
The three polaritons
of each set are thus exactly phase-matched in both energy and
wavevector spaces.  Additionally, the upper and lower states have
always identical group velocities, while all three waves velocities coincide
for a particular value of detuning $k=k_V$ (orange arrows in
Fig.~\ref{fig:transitions-q}).
This might lead to improved non-linear optical response, like in the
optical parametric oscillation
phenomenon,\cite{Shen:2003}
which has been studied in the literature in monolithic semiconductor
  microcavities exploiting exciton 
polaritons\cite{Savvidis:2000,Ciuti:2001,Saba:2001,Ciuti:2003} and
coupled microcavities.\cite{Diederichs:2006}
After pumping on the
central state, entangled photon pairs (idler and signal) would be
expected from the upper and lower branches.  These latter would
propagate along well-defined directions with respect to the pump beam,
allowing an angular discrimination of the beams at the sample outcome
(see below).
Moreover, the generation would be polychromatic (even if possibly
enhanced for $k = k_V$) above the frequency of separation between the two
ISB bare transitions, with angular separation of colors in free space.

\begin{figure}[t]
\includegraphics{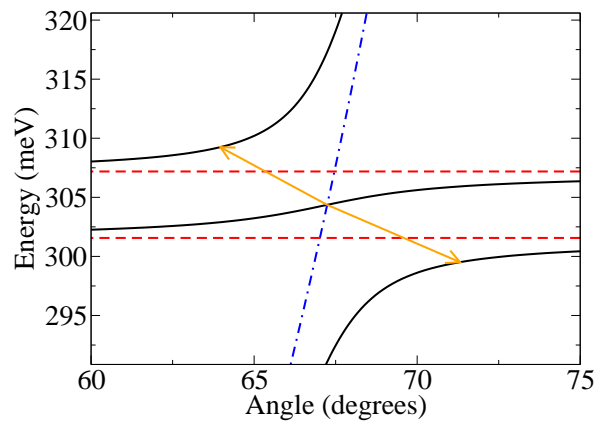}%
\caption{(Color online) Polariton branches for 
  $B=B_2=5.17$~T as a function of the angle of the coupled light
  in the substrate.  The 
  blue dotted--dashed line (red dashed lines) represent the cavity
  mode ($j=0$ and $j=1$ ISB transitions) in absence of
  coupling. The orange arrows connect the same
states as in Fig.~\ref{fig:transitions-q}. Same
  parameters of Fig.~\ref{fig:transitions-B}. 
\label{fig:transitions-theta}}
\end{figure}

Of course, the present model is valid when the typical
  lifetimes of the cavity mode and of the electronic excitations are
  large enough to treat the different transitions independently. For
  what concerns the latter one, calculations performed in a similar
  system\cite{Faugeras:2006} show that the shortest lifetime (due to
  inelastic optical-phonon scattering) is larger than 1.5~ps,
  corresponding to a broadening of about 0.5~meV. The cavity mode
  lifetime can be tuned by tailoring the optical cavity;
  in typical experiments the cavity mode broadening is of the
  same order of the electronic
  excitations one.\cite{Dini:2003} The different states in the
  intermediate and high-field regions are thus expected to be
  distinguishable in the experiments. Additionally, it is also worth
  recalling that in a typical reflectance spectroscopy measurement,
  photons with fixed $(E,k_\parallel)$ propagate in the substrate (of
  index $n_{\text{sub}}$) with fixed angle $\theta$ (with respect to
  the layers normal) given by: 
$$
\sin\theta = \frac{\hbar c}{E \, n_{\text{sub}}} k_\parallel.
$$ 

We show in Fig.~\ref{fig:transitions-theta} the energy versus $\theta$
plot for the polaritonic states around the resonance region of
Fig.~\ref{fig:transitions-q} ($n_{\text{sub}}=3.51$, as in the
experiments of Ref.~\onlinecite{Anappara:2007}). The orange arrows connect the same
states as in Fig.~\ref{fig:transitions-q}, i.e., those for which the
central and generated polaritons have the same group velocities.  As
we can see, the angle difference is small (to ensure they all fall in
the experimental light cone)\cite{Todorov:2008} but sizeable
(slightly less than 10 degrees), so that all three states, even though
broadened, couple to an external mode and can thus be in principle
revealed in an experiment.  

The efficiency of the aforementioned non-linear process relies on
inter-polariton interactions.  In the context of exciton polaritons,
polariton--polariton couplings have been previously
  studied.\cite{Combescot:2007,Glazov:2009,Vladimirova:2010} The corresponding
  scattering matrix elements 
  depend however on the peculiar properties of the exciton components
  of the polaritons. The study of the scattering processes for
  intersubband polaritons in the presence of non-parabolic dispersions
  is however beyond the scope of this work. 

\section{Conclusion}
In conclusion, we have shown that for intersubband polaritons in
narrow-gap semiconductors, with a significant non-parabolicity,
the magnetic field plays  a true
role of an external control parameter that allows to tune the
regime of light--matter interactions. It becomes then possible to tune
the strength of the coupling of the light
with the different non-degenerate intersubband levels. We have
reported numerical results for a InAs/AlSb system, and we have
identified three different regimes for the 
polariton coupling as a function of the intensity of the magnetic field.
Finally, we have presented a design for an optical parametric
oscillator in the FIR spectral range.  The structure is based on the
existence of a mirror dispersion scheme for the magneto-polaritons,
which ideally allows fulfilling phase-matching requirements for the
pump and parametric waves.   

\begin{acknowledgments}
One of the authors (GP) gratefully acknowledges financial support from
Scuola Normale Superiore.
\end{acknowledgments}

\end{document}